\documentclass{article}
\usepackage{authblk}
\usepackage{amsmath}
\usepackage{amssymb}
\usepackage{mathrsfs}
\usepackage{bm}
\usepackage{float}
\usepackage{graphicx}
\usepackage{wrapfig}
\usepackage{url}
\newtheorem{theo}{Theorem}[section]
\title{A liability tracking approach to long term management of pension funds}
\date{\today}
\author[$\dag$]{Masashi Ieda\thanks{ieda@craft.titech.ac.jp}}
\author[$\ddag$]{Takashi Yamashita}
\author[$\dag$]{Yumiharu Nakano}
\affil[$\dag$]{
	Graduate School of Innovation Management \authorcr
	Tokyo Institute of Technology \authorcr
	2-12-1 Ookayama, Meguro-ku, Tokyo, Japan
}
\affil[$\ddag$]{
The Government Pension Investment Fund, Japan \authorcr
1-4-1 Kasumigaseki, Chiyoda-ku, Tokyo Japan
}

\begin{document}

\maketitle

\begin{abstract}
We propose a long term portfolio management method which takes into account a liability.
Our approach is based on the LQG  (Linear, Quadratic cost, Gaussian) control problem framework
 and then the optimal portfolio strategy hedges the liability by directly tracking a benchmark process which represents the liability.
Two numerical results using empirical data published by Japanese organizations are served:
 simulations tracking an artificial liability and an estimated liability of Japanese organization.
The latter one demonstrates that our optimal portfolio strategy can hedge his or her liability. 
\end{abstract}

\section{Introduction}

In the management of pension fund, a long term portfolio strategy taking into account a liability is one of the most significant issue.
The main reason is the demographic changes in the developed countries:
if the working-age population is enough to provide for old age, the liability is a minor issue in the portfolio management.
Since the life expectancy have increased in recent decades, it becomes insufficient to provide for old age.
Furthermore the low birth rate continues and drives up this problem for decades.
Thus pension funds face a challenging phase to construct long term portfolio strategies which hedge their liabilities.

A lot of pension funds except a few ones \cite{cpp} 
 determine their portfolio strategies by the traditional single time period mean variance approach which excludes an evaluation of a liability.
Its intuitive criterion attracts managers of pension funds.
However the single time period approach is unsuitable for a long term portfolio management
 in the sense that it is unable to change the strategy excepting the initial time.
The multi time period approach which arrows the change of the strategy has a problem that the computational complexity growths exponentially.
Hence if we employ this approach, we are usually unable to obtain the optimal portfolio strategy in realistic time.

Therefore the aim of this paper is to propose a long term portfolio strategy which 
(i) involves an evaluation of a liability,
(ii) admits changes of the strategy at any time,
and (iii) is obtained in realistic time.
To tackle this problem, we employ the LQG (Linear, Quadratic cost, Gaussian) control problem (see, e.g.,  Fleming and Rishel \cite{fleming1975deterministic}).
The LQG control problem is a class of stochastic control problem
 and is able to provide the control minimizing the mean square error of a benchmark process and a controlled process.
Roughly speaking our tactic is that we compute the optimal portfolio strategy with the benchmark process which represents the liability.
Then we can track the liability by using our optimal portfolio strategy.
Although it is difficult to obtain the solution of stochastic control problem in general,
the LQG control problem has the analytical solution which assures that we are able to obtain the solution in realistic time and thus it meets our purpose.

A continuous time  stochastic control approach is one of the most popular method to obtain the suitable long term portfolio strategy.
The literature about this approach is quite rich.
The papers treating managements of pension funds are, for instance, as follows: 
Deelstra et al. \cite{Deelstra2003} and Giacinto et al. \cite{DiGiacinto2010} discuss the portfolio management for pension funds with a minimum guarantee;
Menoncin and Scaillet \cite{Menoncin2006} and Gerrard et al. \cite{Gerrard2012} deal with the pension scheme including the de-cumulation phase.
Our study is on the cutting edge in the sense that deal with tracking liabilities directly and constructing a suitable long term portfolio at the same time.

The organization of the present paper is as follows.
We introduce continuous time models of assets and a benchmark in Section \ref{sec:model}.
To fit in the LQG control problem, they are defined by the linear stochastic differential equations (SDEs).
We mention that our portfolio strategy is represented by amounts of assets.
In Section \ref{sec:opt} we define a criterion of the investment performance and provide the optimal portfolio strategy explicitly.
Several numerical results are served in Section \ref{sec:num_res}.
Throughout the section the parameters related to the assets are determined by an empirical data provided by the Government Pension Investment Fund, Japan.
The simulation using an artificial data are discussed in Section \ref{sec:sim_art_lia} and this result gives conditions that our optimal portfolio strategy works well.
Section \ref{sec:sim_emp_lia} provides the case study using an empirical estimations published by the Japanese Ministry of Health, Labour and Welfare.
It demonstrates that our strategy is able to hedge the liability well.

\section{Continuous time models of assets and a benchmark}
\label{sec:model}
In this section, we present mathematical models of assets and benchmark.
The market which we are considering consists of only one risk-free asset and $n$-risky assets and we have $m$-benchmark component processes.

Let $(\Omega,\mathcal{F},\{\mathcal{F}_t\}_{t\geq 0},\mathbb{P})$ be a filtered probability space
 $\{W_t\}_{t \geq 0}$ be a $d$-dimensional Brownian motion where $d=n+m$
 and $\mathcal{L}^2(\mu_T \times \mathbb{P})$ be a space of stochastic processes $\left\{ Z_t \right\}_{t\geq 0}$ which satisfy
$$
\mathbb{E} \left[ \int_0^T | Z_t |^2 dt \right] < \infty.
$$
We denote price processes of the risk-free asset, the risky assets and the benchmark component processes by
 $S^0_t$,$S_t =(S^1_t,\cdots ,S^n_t)^*$ and $Y_t = (Y^1_t,\cdots,Y^m_t)^*$ respectively, here the asterisk means transposition.
To fit in the LQG control problem, we assume that $S^0_t$,$S_t$ and $Y_t$ are governed by the following SDEs:
\begin{align}
	&\begin{cases}
		\displaystyle \frac{dS^0_t}{S^0_t} = r(t) dt,\\
		S^0_0 = s_0 \in \mathbb{R},
	\end{cases}
	 \label{eq:dyn_riskfree}\\
	&\begin{cases}
	 	\displaystyle \frac{dS^i_t}{S^i_t} = b^i(t) dt + \sum_{j=1}^d \sigma^{ij}_S(t) dW^j_t \;, \quad i=1,2,\cdots ,n,\\
	 	S^i_0 = s^i_0 \in \mathbb{R},
	\end{cases}
	\label{eq:dyn_risky}\\
	&\begin{cases}
	  \displaystyle dY_t = \left( \alpha(t)Y_t + h(t) \right) dt + \sigma_Y(t) dW_t \;,\\
	  Y_0 = y_0 \in \mathbb{R}^m,
	\end{cases}
	 \label{eq:dyn_liabilities}
\end{align}
where $r: [0,T] \rightarrow \mathbb{R}$, $b: [0,T] \rightarrow \mathbb{R}^n$, $\sigma_S: [0,T] \rightarrow \mathbb{R}^{n\times d}$,
$\alpha: [0,T] \rightarrow \mathbb{R}^{m\times m}$, $h: [0,T] \rightarrow \mathbb{R}^m$ and $\sigma_Y: [0,T] \rightarrow \mathbb{R}^{m \times d}$
 are deterministic continuous functions 
 and $T < \infty$ represents the maturity.
Coefficients $r$, $b^i$ and $\sigma_S$ stand for the risk-free rate and the expected return rate of the $i$-th asset and the volatility.

Let  a class of portfolio strategy $\mathcal{A}$ be the collection of $\mathbb{R}^n$-valued $\mathcal{F}_t$-adapted process $\left\{ u_t \right\}_{0\leq t \leq T}$ which satisfies 
$$
	\mathbb{E} \left[ \int_0^T | u_t |^2 dt \right] < \infty,
$$
$\xi_t \in \mathbb{R}^n$ be the amount of the risky asset held by an investor at time $t$,
and $X_t$ be the value of our portfolio at time $t$.
Then the amount of  the risk-free asset held by the investor is represented by $X_t - \sum_{i=1}^n \xi^i_t$.
Hence $\{ X_t \}_{0\leq t \leq T}$ is governed by
\begin{align}
	\begin{cases}
		\displaystyle dX_t = \sum_{i=1}^n \xi^i_t \frac{dS^i_t}{S^i_t} + \left( X_t- \sum_{i=1}^n\xi^i_t \right)\frac{dS^0_t}{S^0_t},
		  \quad \{\xi_t\}_{0\leq t \leq T} \in \mathcal{A},\\
		X_0 = x_0 = s^0_0 + s_0^* \mathbf{1},
    \end{cases}
    \label{eq:wealth}
\end{align}
where $\mathbf{1} = (1,\cdots,1)^* \in \mathbb{R}^n$. 
To emphasize the initial wealth and the control variable, we may write $X_t=X^{x_0,\xi}_t$.

The solution $X_t$ of the SDE (\ref{eq:wealth}) is given by the following:
\begin{eqnarray*}
X_t &=& e^{\int_0^t r(u) du}x_0 + \int_0^t e^{\int_s^t r(u) du} (b(s)-r(s)\mathbf{1})^*\xi_s ds\\
	&& \quad + \int_0^t e^{\int_s^t r(u) du} \xi_s^* \sigma_S(s) dW_s.
\end{eqnarray*}
Moreover since $r$, $b$, and $\sigma_S$ are continuous functions on $[0,T]$ and $\xi \in \mathcal{A}$,
 $X_t$ is in $\mathcal{L}^2(\mu_T \times \mathbb{P})$:
\begin{eqnarray*}
\mathbb{E}\left[ \int_0^T X_t^2 dt \right] &\leq&
 	\mathbb{E}\left[ \int_0^T  \left( e^{\int_0^t r(u) du} x_0 \right)^2 dt \right]\\
 	&& \; +\mathbb{E}\left[ \int_0^T \left( \int_0^t e^{\int_s^t r(u)du} (b(s)-r(s)\mathbf{1})^* \xi_s ds \right)^2 dt \right]\\
 	&& \; +\mathbb{E}\left[ \int_0^T \left(  \int_0^t e^{\int_s^t r(u)du} \xi_s^* \sigma_S(s) dW_s \right)^2 dt \right]\\
 	&<& T K_0 x_0^2 + T^2 K_1 \mathbb{E}\left[ \int_0^T \xi_s^* \mathbf{1}\mathbf{1}^* \xi_s ds \right]\\
 	&& \; + T K_2 \mathbb{E}\left[ \int_0^T \xi_s^* \mathbf{1}\mathbf{1}^* \xi_s ds \right]\\
 	&<& \infty,
\end{eqnarray*}
where $K_0$, $K_1$ and $K_2$ are constants.

\section{Optimal investment strategy}
\label{sec:opt}
We define the criterion of investment performance $J$ by
\begin{equation}
\label{eq:opt_pro_gen}
J[\xi]  = \mathbb{E} \left[
	\int_0^T \gamma_1 \left( a(t)^*Y_t - X^{x_0,\xi}_t \right)^2 dt + \gamma_2 \left( A^*Y_T - X^{x_0,\xi}_T \right)^2
	\right],\; \xi \in \mathcal{A},
\end{equation}
where $\gamma_1\geq 0$ and $\gamma_2\geq 0$ are constants, $A\in\mathbb{R}^{m}$ is a constant vector,
 and $a:[0,T]\rightarrow\mathbb{R}^{m}$ is a deterministic continuous function.
Hence our investment problem is to find the control $\hat{\xi}$ s.t. $J[\hat{\xi}] \leq J[\xi]$, $\xi \in \mathcal{A}$.
Since the performance criterion is represented by quadratic functions, our investment problem become the LQG control problem.
We determine $a(t)$, $A_T$ and the parameters of $Y_t$ to be able to regard $a(t)^*Y_t$ and $A_T^*Y_T$ as a liability.

The optimal portfolio strategy is represented in the following form:
\begin{theo}
\label{th:opt_stra}
We define the portfolio strategy $\hat{\xi}$ as follows:
\begin{eqnarray}
	\hat{\xi}_t &=&\frac{-1}{2F^{00}(t)}(\sigma_S(t)\sigma_S(t)^*)^{-1}
	\left[ (b(t)-r(t)\mathbf{1}) (2F^{00}(t) X_t+2\tilde{F}^{0*}(t)Y_t + G^0(t) ) \right. \nonumber \\
	&&\qquad\qquad\qquad\qquad\qquad\qquad
	\left. + 2\sigma_S(t)\sigma_Y^*(t)\tilde{F}^{0}(t) \right]
\end{eqnarray}
where $F^{00},G^0:[0,T]\rightarrow\mathbb{R}$ and $\tilde{F}^0:[0,T]\rightarrow\mathbb{R}^m$ are solutions of  following ordinary differential equations (ODEs):
\begin{align}
	&\begin{cases}
		\displaystyle \frac{d}{dt}F^{00}(t) + \gamma_1 + 2r(t)F^{00}(t)\\
			\qquad- (b(t)-r(t)\mathbf{1})^* (\sigma_S(t)\sigma_S(t)^*)^{-1} (b(t)-r(t)\mathbf{1})F^{00}(t) = 0,\\
		F^{00}(T) = \gamma_2,
    \end{cases}
    \label{eq:ode_f00}\\
	&\begin{cases}
		\displaystyle \frac{d}{dt}\tilde{F}^{0}(t) - \gamma_1 a(t) + r(t)\tilde{F}^{0}(t) + \alpha(t)^*\tilde{F}^0(t) \\
			\qquad - (b(t)-r(t)\mathbf{1})^* (\sigma_S(t)\sigma_S(t)^*)^{-1} (b(t)-r(t)\mathbf{1})\tilde{F}^{0}(t) = 0,\\
		\tilde{F}^{0}(T) = -2\gamma_2A,
	\end{cases}
	\label{eq:ode_f0_tilde}\\
	&\begin{cases}
		\displaystyle \frac{d}{dt}G^{0}(t) + r(t)G^{0}(t) + 2h(t)^* \tilde{F}^{0}(t)\\
		\qquad-(b(t)-r(t)\mathbf{1})^* (\sigma_S(t)\sigma_S(t)^*)^{-1} (b(t)-r(t)\mathbf{1})G^{0}(t)\\
		\qquad\qquad -(b(t)-r(t)\mathbf{1})^* (\sigma_S(t)\sigma_S(t)^*)^{-1}  \sigma_S(t)\sigma_Y^*(t)\tilde{F}^{0}(t) = 0,\\
		G^{0}(T) = 0.
	\end{cases}
	\label{eq:ode_g0}
\end{align}

Then $\hat{\xi}$ satisfies $\hat{\xi} \in \mathcal{A}$ and $ J[\hat{\xi}] \leq J[\xi]$, $\xi \in \mathcal{A}$.
\end{theo}
The proof of Theorem \ref{th:opt_stra} is given in the appendix.

We note that $\hat{\xi}_t$ has feedback terms of $X_t$ and $Y_t$.
This implies that our optimal strategy has delays to catch up the the benchmark process $a(t)^*Y_t$.
Hence the preferable situation applying our strategy is the case that $a(t)^*Y_t$ does not fluctuate violently.

\section{Numerical results}
\label{sec:num_res}
We apply our method to an empirical data provided by the Japanese organizations.
This section is divided to two subsections according to the type of liabilities, an artificial liability and
 the liability constructed by the estimations published by the Ministry of Health, Labour and Welfare of Japan.
The former one suggests the situation that our optimal strategy works well
 and the latter one demonstrates that our portfolio strategy is able to hedge the liability.
 
Before we move on the each subsection, we determine the common parameters in following subsections.
The first task is to determine the parameters relating to the benchmark component processes.
They consists of the income of a pension fund $C_t$ and his or her expense $B_t$ and thus $n=2$ and $Y_t=(C_t,B_t)^*$.
We set the parameters constructing the benchmark process as follows:
$$
\gamma_1=\gamma_2=1, \quad a(t) = (-1,1)^*, \quad A=(-1,1)^*.
$$
Hence the benchmark process is $B_t-C_t$ which represents a shortfall of the income
 and then we regard this shortfall as the liability.
To discuss the performance of the strategy, we introduce a hedging error function of the $i$-th sample path $E^i_t$ and its average $\bar{E}_t$ as follows:
$$
	E^i_t = \left| (B_t-C_t) - X_t^i \right|,\quad
	\bar{E}_t = \sum_{i=1}^N \frac{E^i_t}{N},
$$
where $X_t^i$ is the $i$-th sample path of $X_t$ and $N \in \mathbb{N}$ is the number of the sample path.
We set $N=1000$ except as otherwise noted.

The next task is to determine the risk-free rate and the expected return rates and volatilities of risky assets.
We invest the following four assets: indices of the domestic bond, the domestic stock, the foreign bond and the foreign stock;
 we number them sequentially.
According to the estimations of return rate and volatilities by the Government Pension Investment Fund, Japan \cite{rate},
we construct $b(t)$ and $\sigma_S(t)$ as follows: $b^1(t)=3\%$, $b^2(t)=4.8\%$, $b^3(t)=3.5\%$ and $b^4(t)=5.0\%$;
\begin{align}
	\sigma^{ij}_S(t) =
	&\begin{cases}
		 \tilde{\Sigma}_{ij}  \qquad i,j \in \{ 1,\cdots,n \}, \\
		 0 \qquad \mathrm{otherwise},
    \end{cases}
\end{align}
where $\tilde{\Sigma}$ the Cholesky decomposition of $\Sigma$, a variance-covariance matrix of the assets:
$$
\Sigma =
\begin{pmatrix}
	0.00297025&			0.0018189375&		-0.000439488&	-0.0005409125\\
	0.0018189375&		0.04950625&			-0.00777504&		0.0119248875\\
	-0.000439488&		-0.00777504&			0.01806336&		0.01467312\\
	-0.0005409125&		0.0119248875&		0.01467312&		0.03940225\\
\end{pmatrix}.
$$
We choose a money market account as the risk-free asset and we set $r(t)=0.0\%$.
\subsection{Simulation with an artificial liability}
\label{sec:sim_art_lia}
\begin{figure}[H]
	\begin{minipage}[t]{0.47\columnwidth}
	\begin{center}
		\includegraphics[height=5cm]{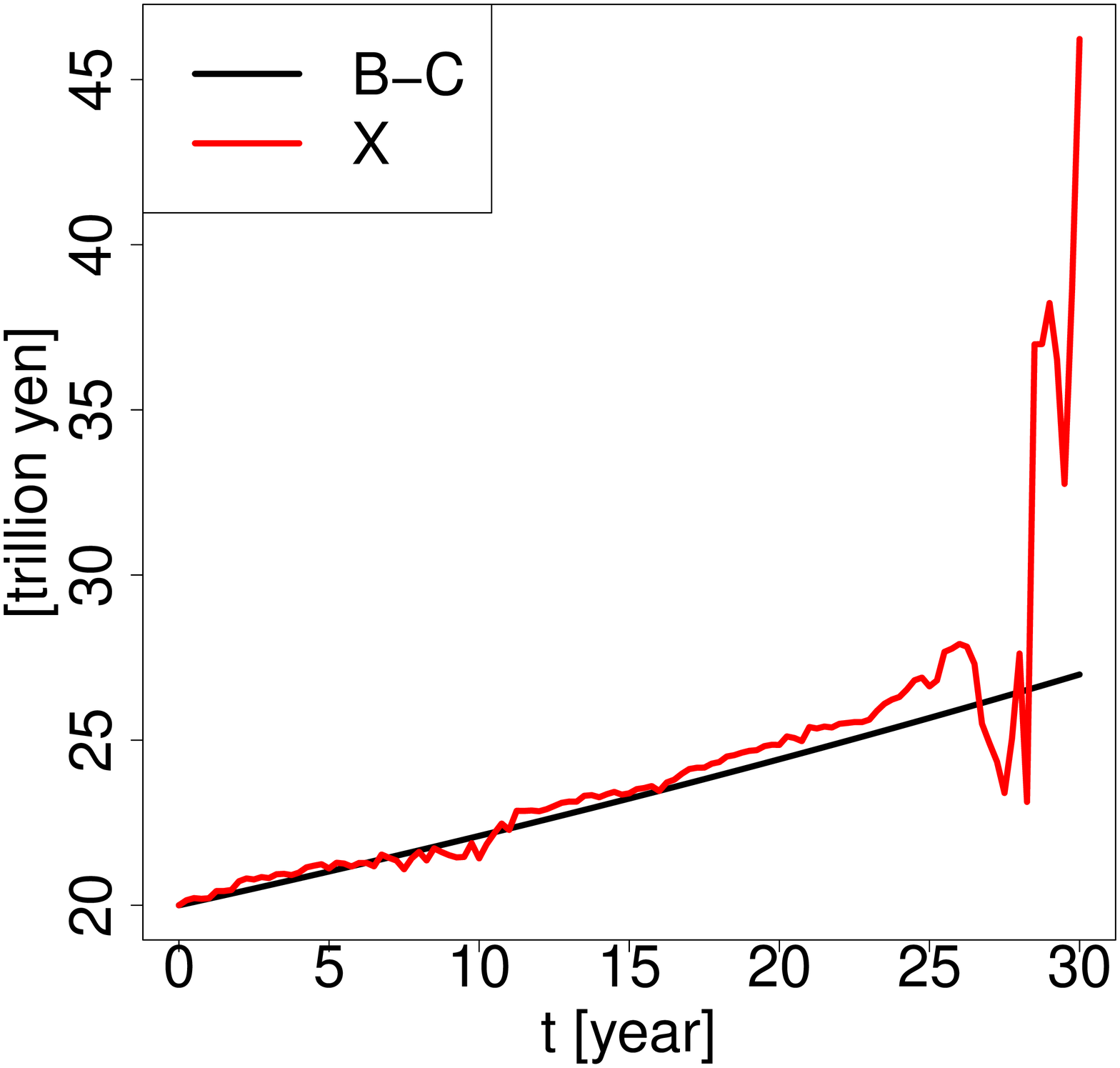}
		\caption{A sample path of $B_t-C_t$ and $X_t$. The black and red lines represent $B_t-C_t$ and $X_t$ respectively.}
		\label{fig:art_30_cbx}
	\end{center}
	\end{minipage}
	\hfill
	\begin{minipage}[t]{0.47\columnwidth}
	\begin{center}
			\includegraphics[height=5cm]{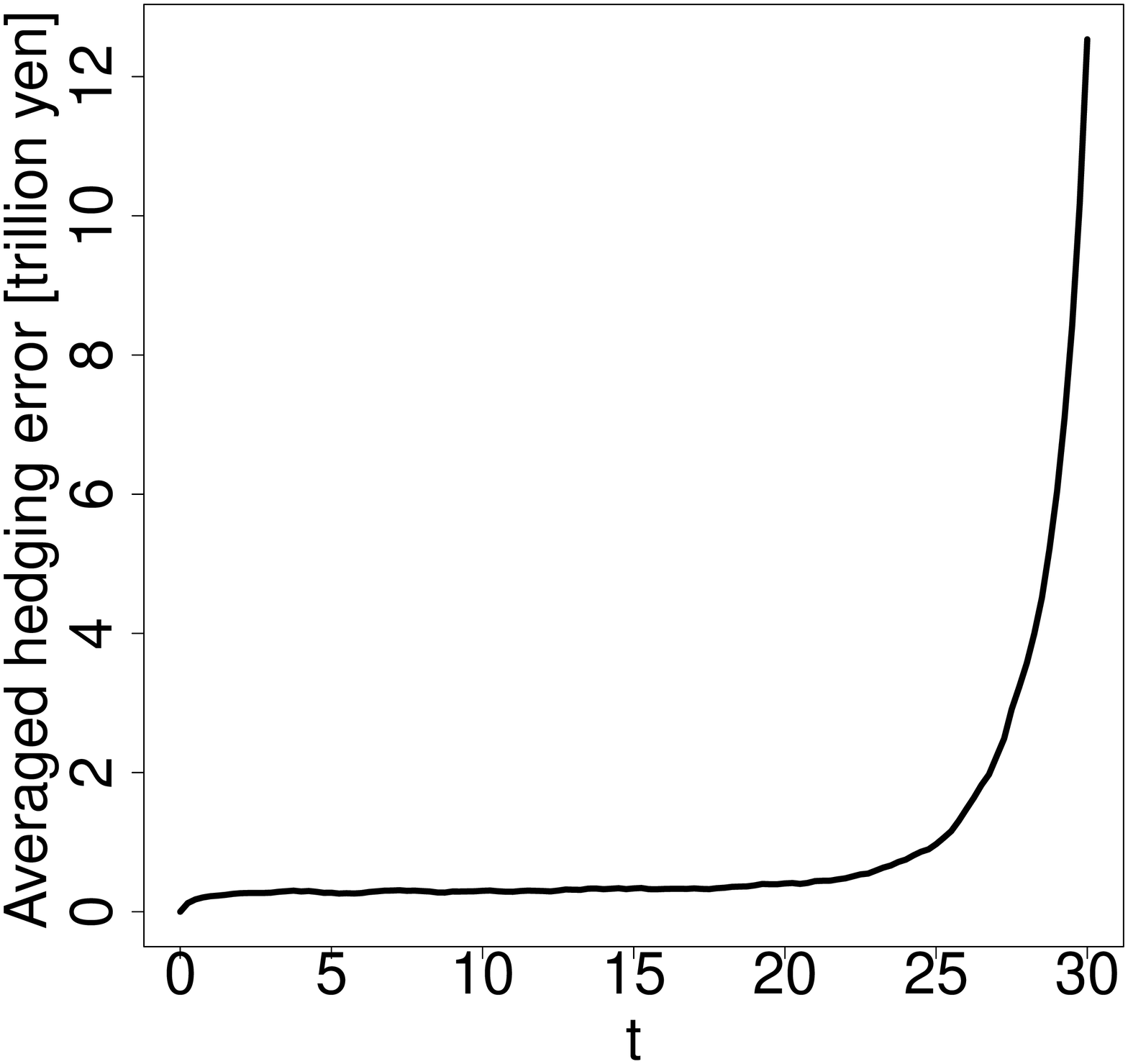}
			\caption{Averaged hedging error $\bar{E}_t$}
			\label{fig:art_30_mean}
	\end{center}
	\end{minipage}
\end{figure}

\begin{figure}[H]
	\begin{center}
		\includegraphics[height=5cm]{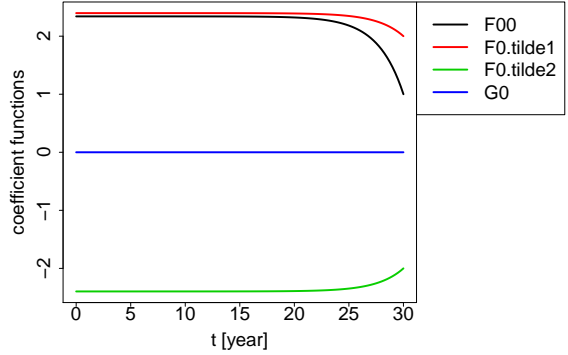}
		\caption{The time evolution of $F^{00}$, $\tilde{F}^0$ and $G^0$.
			The black, red, green and blue lines represent $F^{00}$, $\tilde{F}^{01}$,$\tilde{F}^{02}$ and $G^0$ respectively.}
		\label{fig:art_30_cf}
	\end{center}
\end{figure}

In this subsection we consider the following an artificial deterministic liability model:
\begin{align}
	&\begin{cases}
		 dC_t = 0.01 C_t dt, \\
		 C_0 = 80 \; \mathrm{ [trillion\;yen] },
    \end{cases}\\
    &\begin{cases}
		 dB_t = 0.01 B_t dt, \\
		 B_0 = 100 \; \mathrm{ [trillion\;yen] },
    \end{cases}
\end{align}
i.e., we set $\alpha^{ij}(t)=0.01\delta_{ij}$ and $h(t)=0$.
We assume that our wealth coincides with the benchmark at the initial time: $X_0=B_0-C_0$.
We construct the optimal portfolio strategy over three decades, i.e., $T:=30$
 and we determine the functions $F^{00}$, $\tilde{F}^0$ and $G^0$ by solving the ODEs (\ref{eq:ode_f00})-(\ref{eq:ode_g0}) numerically.
Then we simulate $N$ paths of $(S_t,Y_t)$ on $[0,T]$ according to equation (\ref{eq:dyn_riskfree})-(\ref{eq:dyn_liabilities})
 using a standard Euler-Maruyama scheme with time-step $\Delta t = 0.25$.
Figure \ref{fig:art_30_cbx} describes an investment result of a sample path.
The black and red lines in Figure\ref{fig:art_30_cbx} represent $B_t-C_t$ and $X_t$ respectively.
The most significant issue it indicates is that the performance of the strategy is quite poor near the maturity.
Figure \ref{fig:art_30_mean} describing $\bar{E}_t$ implies that this poor performance does not depend on the sample path.
Figure \ref{fig:art_30_cf} suggests a key factor of this phenomenon: 
values of functions $F^{00}$, $\tilde{F}^0$ and $G^0$ change drastically between $t=25$ and $t=30$;
 this time period coincides with the term the hedging error becomes large rapidly.
Figure \ref{fig:art_30_cf} also implies that the existence of the stationary solutions of the ODEs (\ref{eq:ode_f00})-(\ref{eq:ode_g0}).
As described in Figure\ref{fig:art_30_mean}, the strategy relatively works well on the time period
 when the functions $F^{00}$, $\tilde{F}^0$ and $G^0$ reach the stationary state.
Hence the strategy will be improved by using the stationary solutions of the ODEs (\ref{eq:ode_f00})-(\ref{eq:ode_g0}) on entire region.

\begin{figure}[H]
	\begin{center}
			\includegraphics[height=5cm]{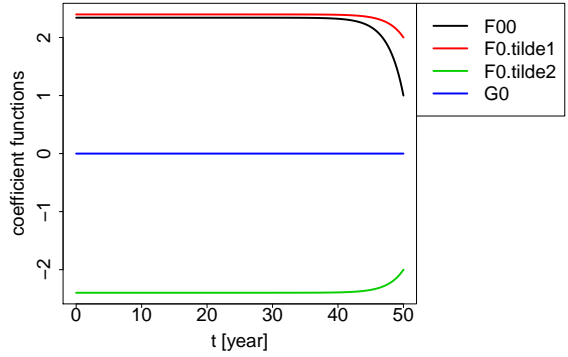}
			\caption{The time evolution of $F^{00}$, $\tilde{F}^0$ and $G^0$ (improved case).
					The black, red, green and blue lines represent $F^{00}$, $\tilde{F}^{01}$,$\tilde{F}^{02}$ and $G^0$ respectively.
					}
			\label{fig:art_50_cf}
	\end{center}
\end{figure}

To obtain the stationary solutions of the ODEs (\ref{eq:ode_f00})-(\ref{eq:ode_g0}), we replace $T$ to a value large enough.
We denote it by $\tilde{T}$ and set $\tilde{T}=50$.
Figure\ref{fig:art_50_cf} shows values of $F^{00}$, $\tilde{F}^0$ and $G^0$ obtained by solving the ODEs (\ref{eq:ode_f00})-(\ref{eq:ode_g0}) with parameter $\tilde{T}$.
We can find that the functions $F^{00}$, $\tilde{F}^0$ and $G^0$ take the stationary solutions on $[0,T]$.
Results of simulations using the improved strategy are described as follows. 
\begin{figure}[H]
	\begin{minipage}[t]{0.47\hsize}
		\includegraphics[height=5cm]{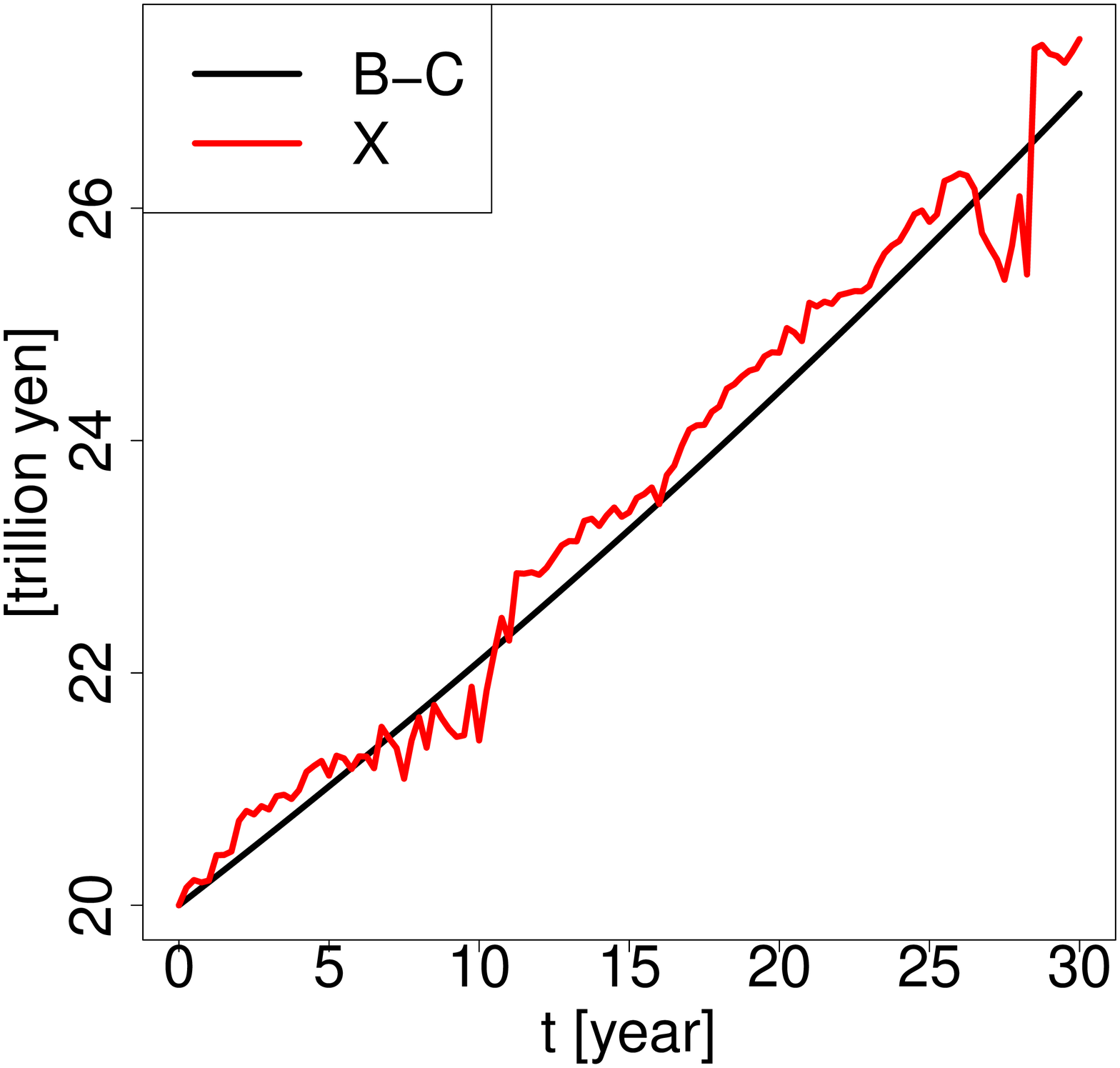}
		\caption{A sample path of $B_t-C_t$ and $X_t$ (improved version). The black and red lines represent $B_t-C_t$ and $X_t$ respectively.}
		\label{fig:art_50_cbx}
	\end{minipage}
	\hfill
	\begin{minipage}[t]{0.47\hsize}
	\begin{center}
			\includegraphics[height=5cm]{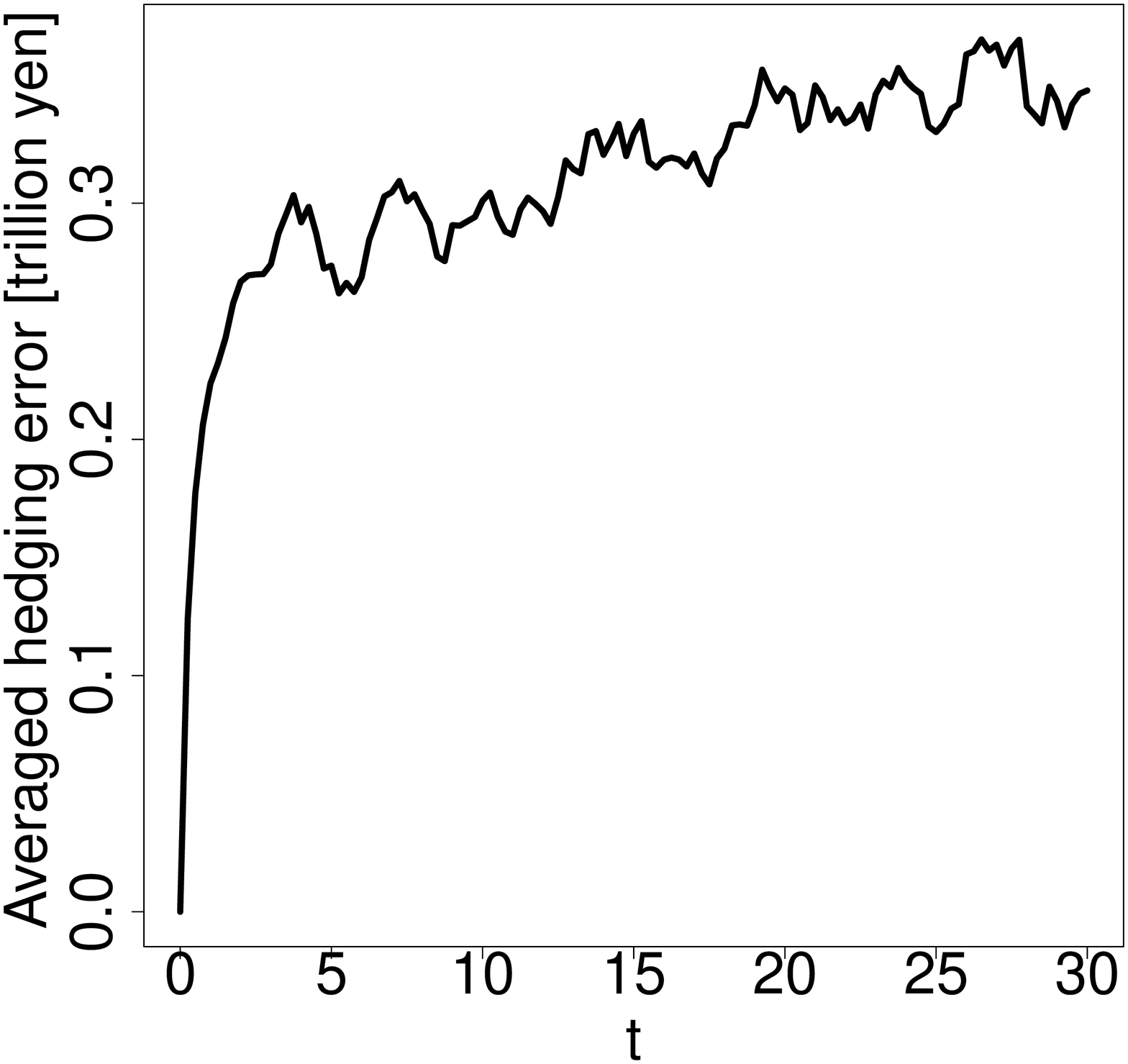}
			\caption{Averaged hedging error $\bar{E}_t$ (improved version).}
			\label{fig:art_50_mean}
	\end{center}
	\end{minipage}
\end{figure}
Figure \ref{fig:art_50_cbx} and Figure \ref{fig:art_50_mean} indicate that the performance near the maturity is improved and it does not depend on the sample paths.
This result leads us to the conclusion that we should construct the strategy with the stationary solutions of the functions $F^{00}$, $\tilde{F}^0$ and $G^0$ if they exists.

\begin{figure}[H]
	\begin{center}
		\includegraphics[height=5cm]{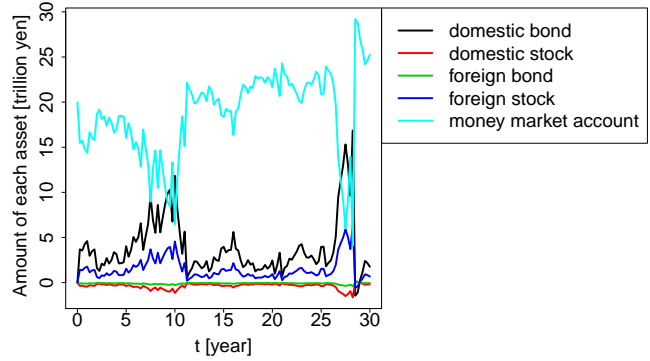}
		\caption{An amount of each asset on the sample path described in Figure \ref{fig:art_50_cbx}.
			The black, red, green, blue and light blue lines represent the amount of a domestic bond, a domestic stock, a foreign bond, a foreign stock and money market account respectively.
		}
		\label{fig:art_50_portfolio}
	\end{center}
\end{figure}

At the end of this subsection, we mention about our portfolio composition.
Figure \ref{fig:art_50_portfolio} displays the asset allocation on the sample path described in Figure \ref{fig:art_50_cbx}.
Money market account, domestic bond and foreign stock indicated by light blue black and blue lines respectively dominate our portfolio.
The optimal strategy is that we keep the most part of the wealth as money market account and compensate for the increment of the benchmark
 by the investment for the domestic bond, low risk and low return asset, and the foreign stock, high risk and high return asset. 
If $X_t$ is deficient in $B_t-C_t$, the strategy increases the proportion of the domestic bond and the foreign stock. 

\subsection{Simulation with an empirical liability}
\label{sec:sim_emp_lia}
\begin{figure}[H]
\label{figProspects}
	\begin{center}
		\includegraphics[height=5cm]{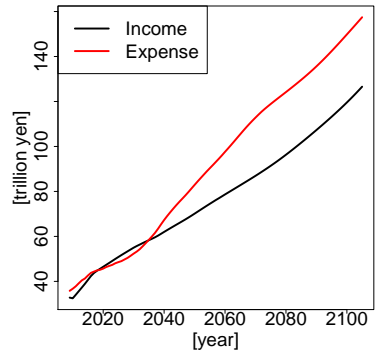}
		\caption{Estimations of income and expense of Japanese national and welfare pensions.
			The black and red lines represent estimations of their income and the expense respectively.}
		\label{fig:prospects}
	\end{center}
\end{figure}

According to the Japanese actuarial valuation published in 2009 \cite{prospect},
 estimated income and expense summed up the national pension and the welfare pension are showed in the figure \ref{fig:prospects}.
We regard these estimations as $C_t$ and $B_t$ and simulate the three decades investments using our optimal strategy
 from 2040 when the shortfall of the pension fund starts to expand drastically.
The following reasons support that this situation is a valid case study:
(i) a phase expanding $B_t-C_t$, the shortfall of the pension fund, is the most typical one expressing the demographic changes;
(ii) the behaviour of $B_t-C_t$ in this term meets the condition to apply our optimal strategy: $B_t-C_t$ is increasing in the entire region.
Throughout this subsection we set the start point as the year 2040, i.e.,  $t=0$ and $t=15$ represent the year 2040 and the year 2055 respectively.

To construct the optimal strategy, we first calibrate $\alpha(t)$, $\sigma_Y(t)$ and $h(t)$ to fit the estimations.
Setting $\alpha(t)=\sigma_Y(t)=0$ and $h(t)$ as a numerical differentiation of the estimations is a simple method to accomplish the purpose.
Since we are discussing the three decades portfolio, we determine $T=30$.
As suggested in Section \ref{sec:sim_art_lia}, we set $\tilde{T}=50$ to obtain the stationary $F^{00}$ and $\tilde{F}^0$.
We are unable to expect the stationary $G^0$ because $h(t)$ explicitly depends on $t$.
We assume that our wealth coincide with the benchmark at the initial time: $X_0=B_0-C_0$.
Then we simulate $N$ paths of $(S_t,Y_t)$ on $[0,T]$ according to equation (\ref{eq:dyn_riskfree})-(\ref{eq:dyn_liabilities})
 using a standard Euler-Maruyama scheme with time-step $\Delta t = 0.25$ which means that we can rearrange our portfolio every quarter.
Results of the simulation are as follows.
\begin{figure}[H]
	\begin{minipage}[t]{0.47\columnwidth}
		\includegraphics[height=5cm]{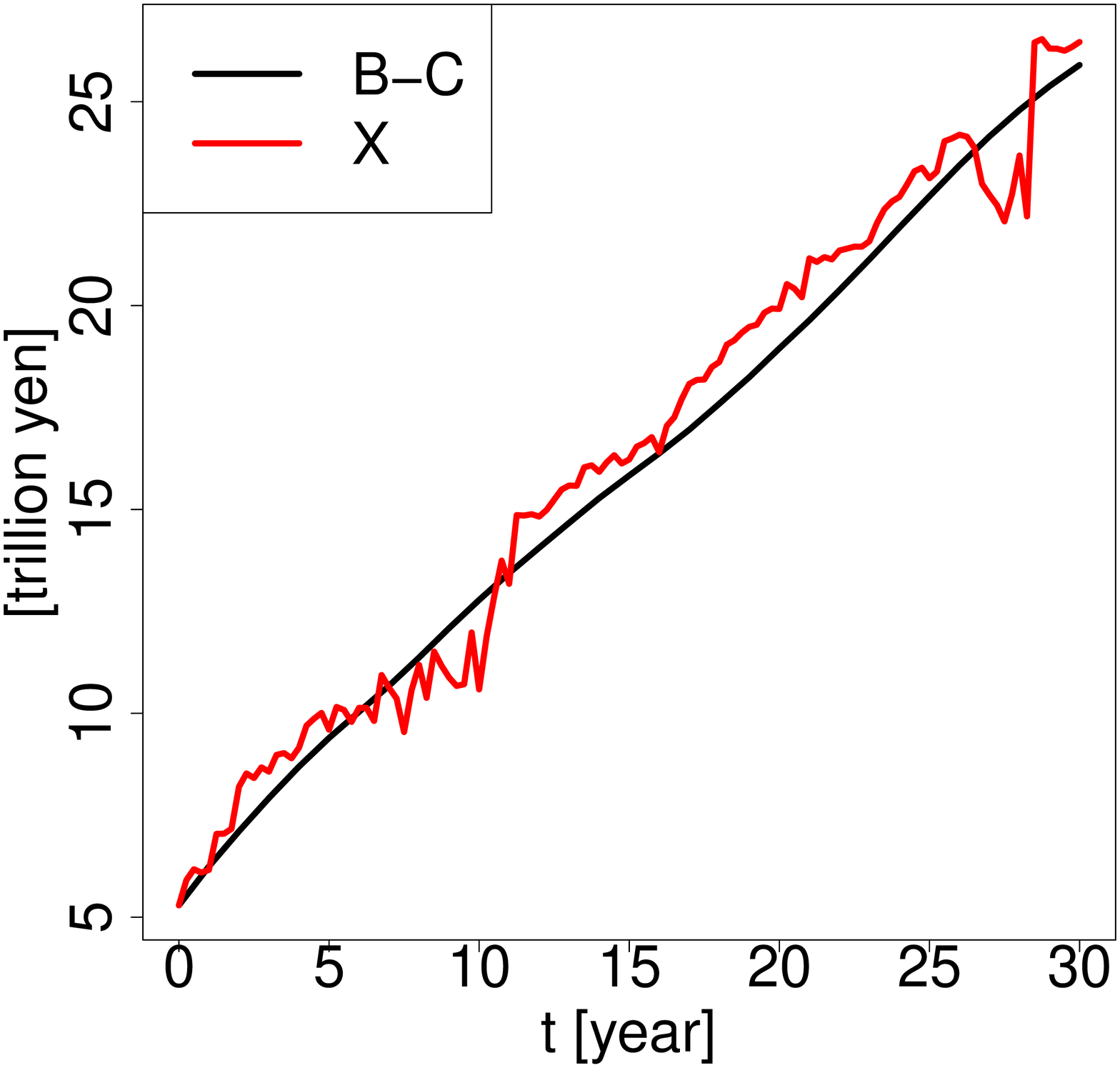}
		\caption{A sample path of $B_t-C_t$ and $X_t$. The black and red lines represent $B_t-C_t$ and $X_t$ respectively.}
		\label{fig:emp_cbx}
	\end{minipage}
	\hfill
	\begin{minipage}[t]{0.47\columnwidth}
	\begin{center}
			\includegraphics[height=5cm]{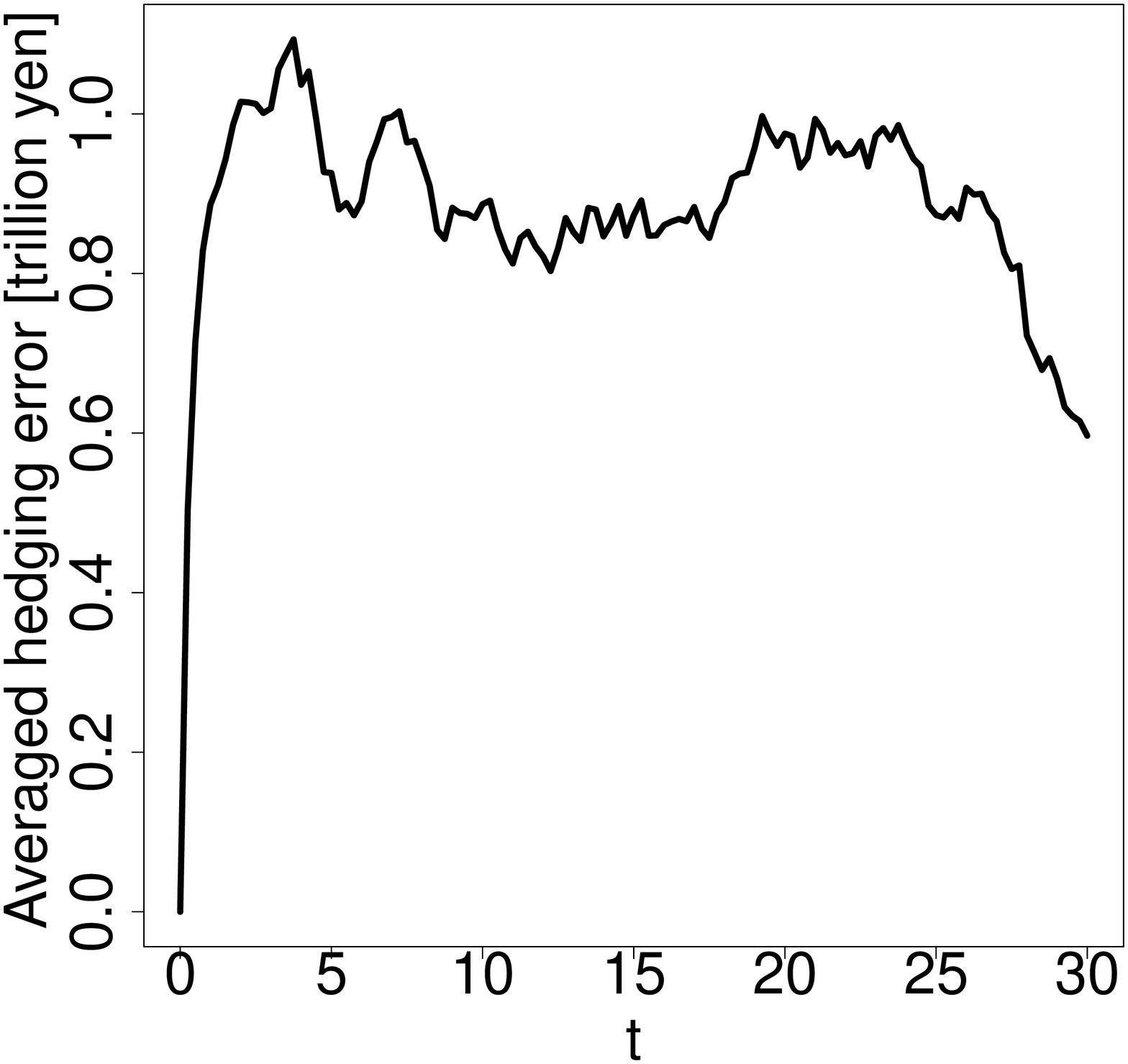}
			\caption{Averaged hedging error $\bar{E}_t$.}
			\label{fig:emp_mean}
	\end{center}
	\end{minipage}
\end{figure}
We are able to argue that our strategy hedges the shortfall well since Figure \ref{fig:emp_mean} suggest that $\bar{E}_t$, the averaged hedging error,
 is approximately 3\% of $B_t-C_t$, the shortfall, in every quarter.
 
\begin{figure}[H]
	\begin{center}
		\includegraphics[height=5cm]{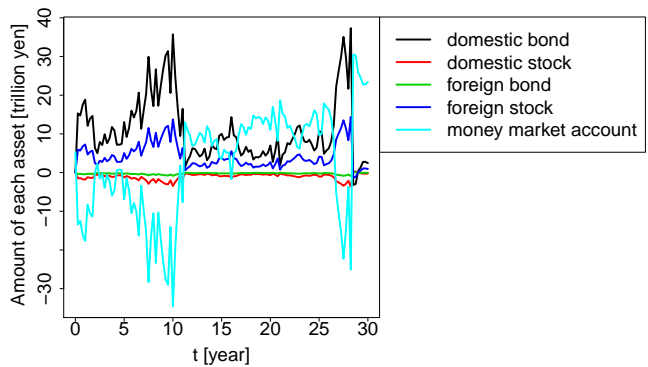}
		\caption{An amount of each asset on the sample path described in Figure \ref{fig:emp_cbx}.
			The black, red, green, blue and light blue lines represent the amount of a domestic bond, a domestic stock, a foreign bond, a foreign stock and money market account respectively.
		}
		\label{fig:emp_portfolio}
	\end{center}
\end{figure}
Figure \ref{fig:emp_portfolio} displays the asset allocation on the sample path described in Figure \ref{fig:emp_cbx}.
In the same manner as in the case of the artificial liabilities discussed in Section \ref{sec:sim_art_lia},
 our optimal portfolio is dominated by the money market account, the domestic bond and the foreign stock.
However the proportion of the domestic bond and the foreign stock is much higher.
We can understand this phenomenon intuitively:
 since the shortfall increases more rapid than that discussed in Section \ref{sec:sim_art_lia},
 the hedging portfolio is rearranged to become more profitable.
The practical suggestion from this fact is that we have to take a risk to track the increasing liability and this is quite natural.

\section{Summary}
We have proposed a long term portfolio management method which takes into account a liability.
The LQG control approach allows us to construct a more suitable long term portfolio strategy than
 myopic one obtained by the single time period mean variance approach
 in the sense that we are able to change the strategy at any time.
Our optimal portfolio strategy hedges the liability by directly tracking the benchmark process which represents the liability.
The strategy is evaluated by the mean square error from the benchmark and hence it is intuitive.
Two numerical simulations are served: the former one suggests the situation that our portfolio strategy works well;
the latter one provide the result with the empirical data published by Japanese organizations.
The result demonstrates that our portfolio strategy is able to hedge the liability,
 the shortfall of the income summed up the national pension and the welfare pension, over three decades.

This study leaves ample scope for further research.
Since our criterion is the mean square error, our portfolio strategy inhibits that our wealth exceeds a liability.
This is the similar problem with the traditional mean variance approach.
One of approaches to avoid it is that we extend criterion which is able to hedge only the case that our wealth goes under the liability.
Then we again face the problem of computability as mentioned in the introductory section.

\section*{Acknowledgment}
This work is partially supported by a collaboration research project with the Government Pension Investment Fund, Japan in 2011-2012.

\appendix
\section{Proof of Theorem \ref{th:opt_stra}}
The value function corresponding to our problem (\ref{eq:opt_pro_gen}) is defined by
\begin{eqnarray*}
V_t(x,y) &=& \inf_{\xi \in \mathcal{A}}\mathbb{E} \left[ \int_t^T \gamma_1 \left( a(s)^*Y_s - X^{x_0,\xi}_s \right)^2 ds \right. \\
	 && \qquad \qquad \left. + \gamma_2 \left( A^*Y_T - X^{x_0,\xi}_T \right)^2 \left| X_t^{x_0,\xi}=x, Y_t=y \right.\right].
\end{eqnarray*}
Hence the corresponding Hamilton-Jacobi-Bellman (HJB) equation is given by
\begin{equation}
\label{eq:hjb}
\inf_{\xi\in\mathcal{A}} \left\{ \partial_t V_t(x,y) + \mathscr{L}^{\xi} V_t(x,y) + \gamma_1(a(t)^*y-x)^2 \right\}=0,
\end{equation}
with terminal condition $V_T(x,y)=\gamma_2(A^*y-x)^2$, $(x,y)\in \mathbb{R}\times\mathbb{R}^m$,
where $\partial_t$ is partial differential operator with respect to $t$
 and $\mathscr{L}^{\xi}$ is the infinitesimal generator of the process $(X_t,Y_t)$:
\begin{eqnarray*}
\mathscr{L}^{\xi} \phi(x,y) &=& \left( r(t)x + (b(t)-r(t)\mathbf{1})^* \xi_t \right) \partial_x \phi(x,y) + \left( \alpha(t)y+h(t) \right)^* \partial_y \phi(x,y)\\
&&\quad + \frac{1}{2} \left[ \xi_t^* \sigma_S(t)\sigma_S(t)^* \xi_t \partial_x^2\phi(x,y) + 2\xi_t^* \sigma_S(t)\sigma_Y(t)^* \partial_y\partial_x\phi(x,y) \right. \\
&& \qquad \qquad \left. \mathrm{Tr} \left( \sigma_Y(t)\sigma_Y(t)^*\partial_y^2 \phi(x,y) \right) \right]
\end{eqnarray*}
for $\phi \in C^2(\mathbb{R}\times\mathbb{R}^m)$.
Here $\partial_x^j$ and $\partial_y^j$ are the $j$-th order partial differential operators with respect to $x$ and $y$.
As $\sigma_S(t)\sigma_S(t)^*$ is positive definite, the infimum of (\ref{eq:hjb}) is attained at
\begin{eqnarray*}
\hat{\xi}_t &=& \frac{-1}{\partial^2_x V_t(x,y)} \left( \sigma_S(t)\sigma_S(t)^* \right)^{-1} \left[ 
	(b(t)-r(t)\mathbf{1}) \partial_xV_t(x,y) \right. \\ 
	&& \qquad\qquad\qquad\qquad\qquad\qquad \left.+ \sigma_S(t)\sigma_Y(t)^* \partial_y\partial_x V_t(x,y) \right],
\end{eqnarray*}
and hence (\ref{eq:hjb}) can be written as
\begin{eqnarray*}
&&\partial_t V_t(x,y) +r(t)x \partial_x V_t(x,y) +\left( \alpha(t)y+h(t) \right)^*\partial_y V_t(x,y) \\
&&\qquad	- \frac{1}{2\partial^2_x V_t(x,y)} (b(t)-r(t)\mathbf{1})^* (\sigma_S(t)\sigma_S(t)^*)^{-1} (b(t)-r(t)\mathbf{1}) \left( \partial_x V_t(x,y) \right)^2 \\
&&\qquad	- \frac{\partial_x V_t(x,y)}{\partial^2_x V_t(x,y)} (b(t)-r(t)\mathbf{1})^* (\sigma_S(t)\sigma_S(t)^*)^{-1} \sigma_S(t)\sigma_Y(t)^* \partial_y \partial_x V_t(x,y) \\
&&\qquad	- \frac{1}{2\partial^2_x V_t(x,y)} \left( \partial_y \partial_x V_t(x,y) \right)^* \sigma_Y(t)\sigma_S(t)^*\sigma_S(t)\sigma_Y(t)^* \partial_y \partial_x V_t(x,y)\\
&&\qquad	+ c \left( x^2 -2xa(t)^*y +y^*a(t)a(t)^*y \right)\\
&&=0.
\end{eqnarray*}
Let us try a value function of the form
$$
V_t(x,y) = F^{00}(t)x^2 + 2x\tilde{F}^0(t)^* y + y^*\tilde{F}(t)y + G^0(t)x + \tilde{G}(t)^*y +g(t),
$$
where $ F^{00},G^0,g:[0,T]\rightarrow \mathbb{R}$, $\tilde{F}^0,\tilde{G}:[0,T] \rightarrow \mathbb{R}^m$ and $\tilde{F}$ is a time-dependant symmetric matrix.
It is straightforward to see that
\begin{align}
	\begin{cases}
		\displaystyle \frac{d}{dt}F^{00}(t) + c + 2r(t)F^{00}(t)\\
			\qquad - (b(t)-r(t)\mathbf{1})^* (\sigma_S(t)\sigma_S(t)^*)^{-1} (b(t)-r(t)\mathbf{1})F^{00}(t) = 0,\\
		F^{00}(T) = C,
	\end{cases}
	\label{eq:app_f00}
\end{align}
\begin{align}
	\begin{cases}
		\displaystyle \frac{d}{dt}\tilde{F}^{0}(t) - ca(t) + r(t)\tilde{F}^{0}(t) + \alpha(t)^*\tilde{F}^0(t)\\
			\qquad - (b(t)-r(t)\mathbf{1})^* (\sigma_S(t)\sigma_S(t)^*)^{-1} (b(t)-r(t)\mathbf{1})\tilde{F}^{0}(t) = 0,\\
		\tilde{F}^{0}(T) = -2CA ,
	\end{cases}
	\label{eq:app_tilde_f0}
\end{align}
\begin{align}
	\begin{cases}
		\displaystyle \frac{d}{dt}G^{0}(t) + r(t)G^{0}(t) + 2h(t)^* \tilde{F}^{0}(t)\\
			\qquad -(b(t)-r(t)\mathbf{1})^* (\sigma_S(t)\sigma_S(t)^*)^{-1} (b(t)-r(t)\mathbf{1})G^{0}(t)\\
		\qquad -(b(t)-r(t)\mathbf{1})^* (\sigma_S(t)\sigma_S(t)^*)^{-1}  \sigma_S(t)\sigma_Y^*(t)\tilde{F}^{0}(t) = 0,\\
		G^{0}(T) = 0 ,
	\end{cases}
	\label{eq:app_g0}
\end{align}
\begin{align}
	\begin{cases}
		\displaystyle \frac{d}{dt}\tilde{F}(t) + ca(t)a(t)^* + 2\alpha(t)^*\tilde{F}(t) - \frac{1}{F^{00}(t)}\tilde{F}^0(t)\tilde{F}^{0}(t)^* = 0,\\
		\tilde{F}(T) = CAA^* ,
	\end{cases}
		\label{eq:app_tilde_f}
\end{align}
\begin{align}
	\begin{cases}
		\displaystyle \frac{d}{dt}\tilde{G}(t) + r(t)\tilde{G}(t)  +\alpha(t)^*\tilde{G}(t) + 2h(t)^* \tilde{F}(t) \\
		\displaystyle	\; - \frac{G^0(t)}{F^{00}(t)} (b(t)-r(t)\mathbf{1})^* (\sigma_S(t)\sigma_S(t)^*)^{-1} (b(t)-r(t)\mathbf{1})\tilde{F}^{0}(t) \\
		\displaystyle	\; -\frac{1}{F^{00}(t)}(b(t)-r(t)\mathbf{1})^* (\sigma_S(t)\sigma_S(t)^*)^{-1}  (\sigma_S(t)\sigma_Y^*(t)\tilde{F}^{0}(t)) \tilde{F}^{0}(t) = 0,\\
		\tilde{G}(T) = 0, 
	\end{cases}
		\label{eq:app_tilde_g}
\end{align}
\begin{align}
	\begin{cases}
		\displaystyle \frac{d}{dt}g(t) + h(t)\tilde{G}(t)  +\mathrm{Tr}(\sigma_Y\sigma_Y^*\tilde{F}(t))\\
			\displaystyle \qquad - \frac{(G^0(t))^2}{4F^{00}(t)} (b(t)-r(t)\mathbf{1})^* (\sigma_S(t)\sigma_S(t)^*)^{-1} (b(t)-r(t)\mathbf{1})\\
			\displaystyle \qquad -\frac{G^0}{2F^{00}(t)}(b(t)-r(t)\mathbf{1})^* (\sigma_S(t)\sigma_S(t)^*)^{-1}  \sigma_S(t)\sigma_Y^*(t)\tilde{F}^{0}(t) = 0,\\
		g(T) = 0,
	\end{cases}
		\label{eq:app_g}
\end{align}
and then the associated candidate strategy is represented by
\begin{eqnarray}
	\hat{\xi}_t &=&\frac{-1}{2F^{00}(t)}(\sigma_S(t)\sigma_S(t)^*)^{-1}
	\left[ (b(t)-r(t)\mathbf{1}) (2F^{00}(t) X_t+2\tilde{F}^{0*}(t)Y_t + G^0(t) ) \right. \nonumber \\
	&&\qquad\qquad\qquad\qquad\qquad\qquad
	\left. + 2\sigma_S(t)\sigma_Y^*(t)\tilde{F}^{0}(t) \right].
\end{eqnarray}
Since (\ref{eq:app_f00}) and (\ref{eq:app_tilde_f0}) are linear ODEs, they have unique solutions on $[0,T]$.
The existence of unique $\tilde{F}^{0}(t)$ suggests that (\ref{eq:app_g0}) and (\ref{eq:app_tilde_f}) are linear ODEs.
Hence (\ref{eq:app_g0}) and (\ref{eq:app_tilde_f}) also have unique solutions on $[0,T]$.
In the same manner, the existence of unique solutions of (\ref{eq:app_tilde_g}) and (\ref{eq:app_g}) are guaranteed.
Therefore $\hat{\xi}_t \in \mathcal{A}$ since $X_t$ and $Y_t$ in $\mathcal{L}^2(\mu_T \times \mathbb{P})$.

We now start the verification, i.e., we show that $V_0(x_0,y_0) \leq J[\xi]$, $\xi \in \mathcal{A}$ and $V_0(x_0,y_0)=J[\hat{\xi}]$.
To this end, we introduce a sequence of stopping times $\left\{ \tau_l \right\}_{l\in \mathbb{N}}$ s.t.
$$
\tau_l = \inf \left\{ u \geq 0 : \int_0^u \left| X_t\xi_t \right|^2 dt \geq l, \; \int_0^u \left| Y_t\xi_t \right|^2 dt \geq l \right\}.
$$ 
Applying the Ito formula to $V_{T \wedge \tau_l}(X_{T \wedge \tau_l},Y_{T \wedge \tau_l})$ and taking expectation, we have
\begin{eqnarray*}
V_0(x_0,y_0) &=& \mathbb{E} \left[ \int_0^{T \wedge \tau_l} \left( -\partial_t V_t(X_t,Y_t) - \mathscr{L}^\xi V_t(X_t,Y_t) \right) dt
	 + V_T(X_{T \wedge \tau_l},Y_{T \wedge \tau_l}) \right]\\
&& \quad - \mathbb{E} \left[ \int_0^{T \wedge \tau_l} \partial_x V_t(X_t,Y_t) \xi^* \sigma_S(t) dW_t  \right]\\
&& \quad - \mathbb{E} \left[ \int_0^{T \wedge \tau_l} \left( \partial_x V_t(X_t,Y_t) \right)^* \sigma_Y(t) dW_t \right].
\end{eqnarray*}
As $X_t$ and $Y_t$ are in $\mathcal{L}^2(\mu_T \times \mathbb{P})$ and $\sigma_Y$, $\tilde{F}^0$, $\tilde{F}$ and $\tilde{G}$ are continuous functions on $[0,T]$,
the last term vanishes:
\begin{eqnarray*}
\mathbb{E} \left[ \int_0^{T \wedge \tau_l} \left(\partial_y V_t(X_t,Y_t) \right)^* \sigma_Y(t) dW_t \right]
 	&=&2\mathbb{E}\left[ \int_0^{T \wedge \tau_l} X_t \tilde{F}^0(t)^* \sigma_Y(t) dW_t \right]\\
	&& \;+2\mathbb{E}\left[ \int_0^{T \wedge \tau_l} Y_t^* \tilde{F}(t) \sigma_Y(t) dW_t \right]\\
	&& \; +\mathbb{E}\left[ \int_0^{T \wedge \tau_l} \tilde{G}(t)^* \sigma_Y(t) dW_t \right]\\
	&=& 0.
\end{eqnarray*}
By the definition of $\tau_l$, the continuity of the functions $F^{00}$, $\tilde{F}^0$ and $G^0$, and the fact that $\xi\in\mathcal{A}$,
 the remaining stochastic integral term also vanishes:
\begin{eqnarray*}
\mathbb{E} \left[ \int_0^{T \wedge \tau_l} \partial_x V_t(X_t,Y_t) \xi_t^* \sigma_S(t) dW_t \right] 
	&=& 2\mathbb{E} \left[ \int_0^{T \wedge \tau_l} F^{00}(t)X_t \xi_t^* \sigma_S(t) dW_t \right]\\
	&& \; + 2\mathbb{E} \left[ \int_0^{T \wedge \tau_l} \tilde{F}^0(t)^*Y_t \xi_t^* \sigma_S(t) dW_t \right]\\
	&& \; + \mathbb{E} \left[ \int_0^{T \wedge \tau_l} G^0(t) \xi_t^* \sigma_S(t) dW_t \right]\\
	&=& 0.
\end{eqnarray*}
Since $\tau_l \nearrow \infty$ when $l$ goes to infinity, we get
$$
V_0(x_0,y_0) = \mathbb{E} \left[ \int_0^{T} \left( -\partial_t V_t(X_t,Y_t) - \mathscr{L}^\xi V_t(X_t,Y_t) \right) dt + V_T(X_{T},Y_{T}) \right].
$$
By the HJB equation (\ref{eq:hjb}) and its terminal condition, we obtain
$$
V_0(x_0,y_0) \leq \mathbb{E} \left[ \int_0^T \gamma_1(a(t)^*y-x)^2 dt + \gamma_2(A^*y-x)^2 \right],
$$
which means that $V_0(x_0,y_0)\leq J[\xi]$, $\xi\in\mathcal{A}$.
In the same manner we find that $V_0(x_0,y_0)=J[\hat{\xi}]$ and then the claim is established.

\bibliography{library,text,gpif}
\end{document}